# Measurement of the $^{235}$U(n,f) cross section relative to the $^6$Li(n,t) and $^{10}$B(n,α) standards from thermal to 170 keV neutron energy range at n_TOF

S. Amaducci[1,35], L. Cosentino[1], M. Barbagallo[2], N. Colonna[2], A. Mengoni[3,4], C. Massimi[4,5], S. Lo Meo[3,4], P. Finocchiaro[1,#], O. Aberle[6], J. Andrzejewski[7], L. Audouin[8], M. Bacak[9,6,10], J. Balibrea[11], F. Bečvář[12], E. Berthoumieux[10], J. Billowes[13], D. Bosnar[14], A. Brown[15], M. Caamaño[16], F. Calviño[17], M. Calviani[6], D. Cano-Ott[11], R. Cardella[6], A. Casanovas[17], F. Cerutti[6], Y. H. Chen[8], E. Chiaveri[6,13,18], G. Cortés[17], M. A. Cortés-Giraldo[18], L. A. Damone[2,19], M. Diakaki[10], C. Domingo-Pardo[20], R. Dressler[21], E. Dupont[10], I. Durán[16], B. Fernández-Domínguez[16], A. Ferrari[6], P. Ferreira[22], V. Furman[23], K. Göbel[24], A. R. García[11], A. Gawlik[7], S. Gilardoni[6], T. Glodariu†[25], I. F. Gonçalves[22], E. González-Romero[11], E. Griesmayer[9], C. Guerrero[18], F. Gunsing[10,6], H. Harada[26], S. Heinitz[21], J. Heyse[27], D. G. Jenkins[15], E. Jericha[9], F. Käppeler[28], Y. Kadi[6], A. Kalamara[29], P. Kavrigin[9], A. Kimura[26], N. Kivel[21], I. Knapova[12], M. Kokkoris[29], M. Krtička[12], D. Kurtulgil[24], E. Leal-Cidoncha[16], C. Lederer[30], H. Leeb[9], J. Lerendegui-Marco[18], S. J. Lonsdale[30], D. Macina[6], A. Manna[4,5], J. Marganiec[7,31], T. Martínez[11], A. Masi[6], P. Mastinu[32], M. Mastromarco[2], E. A. Maugeri[21], A. Mazzone[2,33], E. Mendoza[11], P. M. Milazzo[34], F. Mingrone[6], A. Musumarra[1,35], A. Negret[25], R. Nolte[31], A. Oprea[25], N. Patronis[36], A. Pavlik[37], J. Perkowski[7], I. Porras[38], J. Praena[38], J. M. Quesada[18], D. Radeck[31], T. Rauscher[39,40], R. Reifarth[24], C. Rubbia[6], J. A. Ryan[13], M. Sabaté-Gilarte[6,18], A. Saxena[41], P. Schillebeeckx[27], D. Schumann[21], P. Sedyshev[23], A. G. Smith[13], N. V. Sosnin[13], A. Stamatopoulos[29], G. Tagliente[2], J. L. Tain[20], A. Tarifeño-Saldivia[17], L. Tassan-Got[8], S. Valenta[12], G. Vannini[4,5], V. Variale[2], P. Vaz[22], A. Ventura[4], V. Vlachoudis[6], R. Vlastou[29], A. Wallner[42], S. Warren[13], C. Weiss[9], P. J. Woods[30], T. Wright[13], P. Žugec[14,6]

[1]INFN Laboratori Nazionali del Sud, Catania, Italy
[2]Istituto Nazionale di Fisica Nucleare, Sezione di Bari, Italy
[3]Agenzia nazionale per le nuove tecnologie (ENEA), Bologna, Italy
[4]Istituto Nazionale di Fisica Nucleare, Sezione di Bologna, Italy
[5]Dipartimento di Fisica e Astronomia, Università di Bologna, Italy
[6]European Organization for Nuclear Research (CERN), Switzerland
[7]University of Lodz, Poland
[8]Institut de Physique Nucléaire, CNRS-IN2P3, Univ. Paris-Sud, Université Paris-Saclay, F-91406 Orsay Cedex, France
[9]Technische Universität Wien, Austria
[10]CEA Irfu, Université Paris-Saclay, F-91191 Gif-sur-Yvette, France
[11]Centro de Investigaciones Energéticas Medioambientales y Tecnológicas (CIEMAT), Spain
[12]Charles University, Prague, Czech Republic
[13]University of Manchester, United Kingdom
[14]Department of Physics, Faculty of Science, University of Zagreb, Zagreb, Croatia
[15]University of York, United Kingdom
[16]University of Santiago de Compostela, Spain
[17]Universitat Politècnica de Catalunya, Spain
[18]Universidad de Sevilla, Spain
[19]Dipartimento di Fisica, Università degli Studi di Bari, Italy
[20]Instituto de Física Corpuscular, CSIC - Universidad de Valencia, Spain
[21]Paul Scherrer Institut (PSI), Villingen, Switzerland
[22]Instituto Superior Técnico, Lisbon, Portugal
[23]Joint Institute for Nuclear Research (JINR), Dubna, Russia
[24]Goethe University Frankfurt, Germany
[25]Horia Hulubei National Institute of Physics and Nuclear Engineering, Romania
[26]Japan Atomic Energy Agency (JAEA), Tokai-mura, Japan
[27]European Commission, Joint Research Centre, Geel, Retieseweg 111, B-2440 Geel, Belgium
[28]Karlsruhe Institute of Technology, Campus North, IKP, 76021 Karlsruhe, Germany
[29]National Technical University of Athens, Greece
[30]School of Physics and Astronomy, University of Edinburgh, United Kingdom
[31]Physikalisch-Technische Bundesanstalt (PTB), Bundesallee 100, 38116 Braunschweig, Germany
[32]Istituto Nazionale di Fisica Nucleare, Sezione di Legnaro, Italy
[33]Consiglio Nazionale delle Ricerche, Bari, Italy
[34]Istituto Nazionale di Fisica Nucleare, Sezione di Trieste, Italy
[35]Dipartimento di Fisica e Astronomia, Università di Catania, Italy
[36]University of Ioannina, Greece
[37]University of Vienna, Faculty of Physics, Vienna, Austria

[38] University of Granada, Spain
[39] Department of Physics, University of Basel, Switzerland
[40] Centre for Astrophysics Research, University of Hertfordshire, United Kingdom
[41] Bhabha Atomic Research Centre (BARC), India
[42] Australian National University, Canberra, Australia

[#] corresponding author, email: finocchiaro@lns.infn.it

**Abstract**

The $^{235}$U(n,f) cross section was measured in a wide energy range at n_TOF relative to $^6$Li(n,t) and $^{10}$B(n,$\alpha$), with high resolution and in a wide energy range, with a setup based on a stack of six samples and six silicon detectors placed in the neutron beam. This allowed us to make a direct comparison of the reaction yields under the same experimental conditions, and taking into account the forward/backward emission asymmetry. A hint of an anomaly in the 10÷30 keV neutron energy range had been previously observed in other experiments, indicating a cross section systematically lower by several percent relative to major evaluations. The present results indicate that the evaluated cross section in the 9÷18 keV neutron energy range is indeed overestimated, both in the recent updates of ENDF/B-VIII.0 and of the IAEA reference data. Furthermore, these new high-resolution data confirm the existence of resonance-like structures in the keV neutron energy region. The new, high accuracy results here reported may lead to a reduction of the uncertainty in the 1÷100 keV neutron energy region. Finally, the present data provide additional confidence on the recently re-evaluated cross section integral between 7.8 and 11 eV.

## 1 Introduction

The $^{235}$U(n,f) cross section is one of the most important and widely used cross sections. Although it is a standard at the 0.025 eV thermal neutron energy point and between 0.15 and 200 MeV, it is used as reference at all energies for a variety of purposes, such as for the measurement of the neutron fluence for various applications, or for the measurement of the fission cross section of other actinides. Together with other reactions, the neutron-induced fission of $^{235}$U is routinely used at the n_TOF facility at CERN for the neutron beam characterization. A recent high-accuracy determination of the n_TOF neutron flux in the first experimental area (EAR1) for Phase-II, covering the years 2009-2011 [1], made use of four independent detection systems based on three different neutron converting reactions. The Silicon-based SiMon device [2], relying on the $^6$Li(n,t)$^4$He converting reaction, was used to cover the neutron energy range between thermal and 100 keV. The same range was also covered by a MicroMegas detector exploiting the $^{10}$B(n,$\alpha$)$^7$Li reaction [3]. For the higher energy range the $^{235}$U(n,f) reaction was exploited, with different fission fragment detectors: the MicroMegas, a calibrated ionization chamber from PTB [4] and finally Parallel Plate Avalanche Counters up to 1 GeV neutron energy range [5]. After properly normalizing the data to the respective standard cross sections at thermal energy, the flux extracted on the basis of the $^6$Li(n,t)$^4$He and $^{10}$B(n,$\alpha$)$^7$Li reactions mostly agreed with each other and with the results of FLUKA simulations of the neutron beam [6]. Conversely, the flux extracted on the basis of the $^{235}$U(n,f) reaction, determined independently with the PTB and Micromegas detectors, resulted several percent lower in the energy range 10÷30 keV [1]. A possible explanation was that the evaluated fission cross section in that neutron energy region was overestimated by several percent. Although the $^{235}$U(n,f) cross section in that range is not a standard such a large difference was unexpected, in particular since most evaluations available at that time quoted an uncertainty on this cross section below or around 1% (see for example the compilation of standard cross sections in Refs. [7] and [8]).

The evaluated $^{235}$U fission cross section (in ENDF/B-VII.1 [9] and more recently in ENDF/B-VIII.0 [10] and JEFF3.3 [11]) in this neutron energy region relies on the available EXFOR data which mostly date back to the 70's and 80's [12]÷[18]. Those data are shown in the region of interest in Figure 1, in comparison with the two more recent evaluations which apparently tend to overestimate most datasets.

It is also interesting to notice the presence of structures in the cross section, most probably related to the grouping of unresolved resonances, similar to those determined in a measurement of the Au(n,$\gamma$) cross section at n_TOF [19]. Moreover, in a recent paper [20] Jandel et al. measured the $^{235}$U(n,$\gamma$) capture cross section relative to $^{235}$U(n,f), and found that their data between 10 and 30 keV are about 10% larger than the corresponding data from the evaluations in ENDF/B-VII.1 and JENDL-4.0 [21]. While they concluded that the problem was related to the capture cross section, the observed discrepancy could be at least partially attributed to an overestimate of the evaluated fission cross section used as reference.

Such a possible difference in cross section, which has a negligible influence on thermal reactors, can be important for future fast critical or subcritical reactors. Furthermore, the interest in the $^{235}$U(n,f) reaction is more general as it is often used at that energy to determine the neutron flux, or as a reference in measurements of fission cross section of other actinides, of interest for transmutation projects as well as of key importance for the correct modeling of the fission recycling in r-process nucleosynthesis [22]. In order to clarify this issue and reduce the uncertainty in this energy region, a high-accuracy high-resolution measurement of the $^{235}$U(n,f) cross section was performed at n_TOF in EAR1, relative to two cross section standards commonly used as reference. Although data were collected in a wide energy range from thermal neutron energy to 170 keV, we focus here on the unresolved resonance region (from ≈ 2 keV upward), while the detailed analysis of the resonance region will be the subject of a forthcoming paper. The paper is organized as follows: in Section 2 the experimental setup and the data analysis procedure are described, Section 3 reports the results and a discussion is presented in Section 4, followed by Summary and Conclusions.

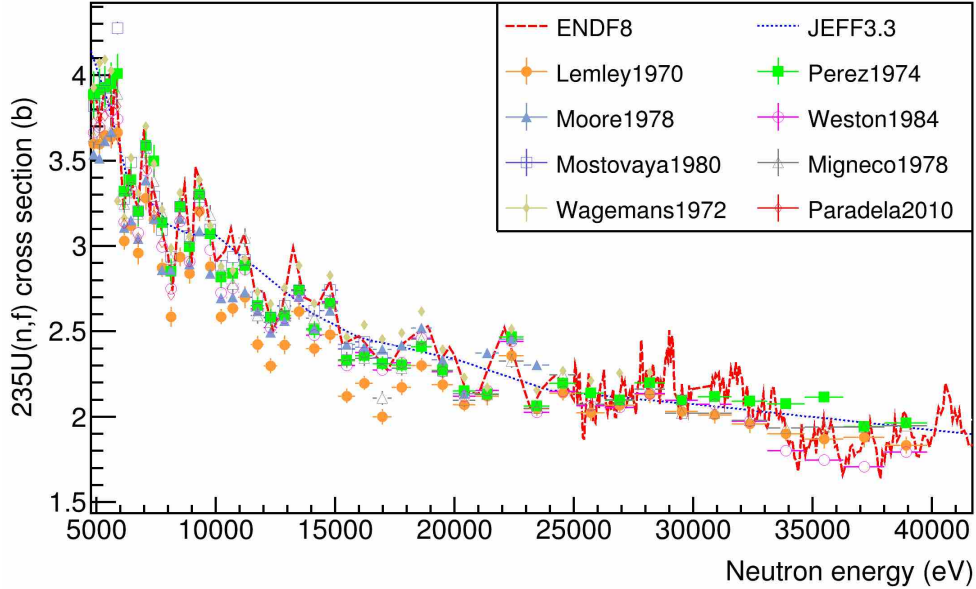

Figure 1 - (Color online) A few $^{235}$U(n,f) cross section datasets from EXFOR in comparison with two evaluations, in the energy range 5÷40 keV.

## 2  Experimental setup

The experiment was performed at n_TOF in EAR1, at the end of a flight path of ≈ 184 m length. In this area, the neutron beam has an instantaneous flux of $10^5 \div 10^6$ n/bunch, an energy ranging from thermal up to ~1 GeV and an energy resolution of $10^{-4}$ up to a few keV. More details on the n_TOF facility and EAR1 can be found in Ref. [23]. The $^{235}$U(n,f) cross section was determined relative to the two main reference reactions, namely $^6$Li(n,t) and $^{10}$B(n,α), whose cross sections are standards of measurement. The reference reactions, their decay products and kinetic energies (for incident thermal neutrons) are listed below.

$$^6Li + n \rightarrow\ ^3H\ (2.73\ MeV) + \alpha\ (2.05\ MeV)$$
$$^{10}B + n \rightarrow\ ^7Li\ (1.01\ MeV) + \alpha\ (1.78\ MeV)\ \text{ground state, BR} \approx 6\%$$
$$^{10}B + n \rightarrow\ ^7Li\ (0.84\ MeV) + \alpha\ (1.47\ MeV) + \gamma\ (0.48\ MeV)\ \text{excited state, BR} \approx 94\%$$

At variance with the $^{235}$U(n,f) cross section, adopted as standard at 0.025 eV and between 150 keV and 200 MeV, the cross section for the two reference reactions are considered standard between 0.025 eV and 1 MeV. The choice of the ratio method (described later in more detail), ensures a minimization of the systematic errors related to the determination of the neutron flux, as well as to geometrical details of the setup and other experimental effects.

The experimental setup consists of a stack of silicon detectors, chosen for their high energy resolution on the reaction products, in particular for the reference reactions. The setup, shown in Figure 2, was mounted in a vacuum chamber inserted in the neutron beam. It consisted of six samples, two for each target material, and six silicon detectors each one facing one sample. Their arrangement was such that each reaction was measured with a separate sample-detector pair in the forward and in the backward direction with respect to beam (it is worth remarking that due to the thickness of samples and substrates, only reaction products exiting from one side are detected). The reason of this arrangement was to introduce a redundancy and compensate (to a large extent) the forward/backward emission asymmetry of the products of the n+$^{10}$B and n+$^6$Li reactions. The six silicon detectors had an active area of 50x50 mm$^2$, and 200 µm thickness, with a 0.5 µm aluminum dead layer on top. The characteristics of the six samples are listed in Table 1. As for the beam shape, the geometrical distribution of the neutrons in the transverse plane is gaussian with a standard deviation of 7 mm.

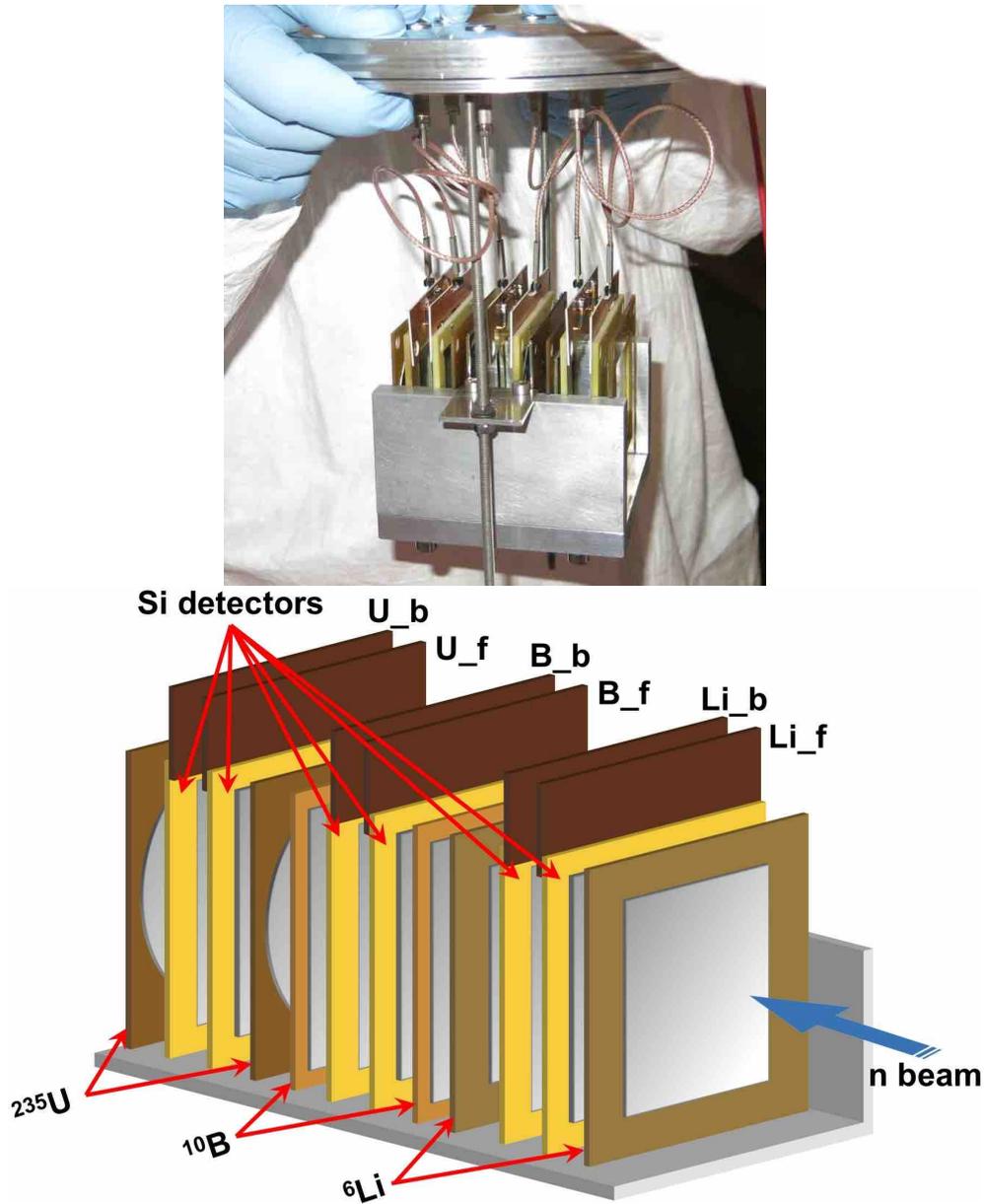

Figure 2 – (Color online) The experimental setup, with the six targets and six silicon detectors used in the measurement.

Table 1 – Characteristics of the six samples (two for each type of deposit)

| Deposit | Size [mm] | Nominal thickness [μm] | Backing | Isotope | Enrichment | Atoms/cm$^2$ | Preparation method |
|---|---|---|---|---|---|---|---|
| LiF | 47 x 47 | 1.97 | Al 50 μm | $^6$Li | 95% | $1.14 \cdot 10^{19}$ | evaporation |
| B$_4$C | 70 x 70 | 0.08 | Al 18 μm | $^{10}$B | 99% | $8.28 \cdot 10^{17}$ | plasma deposition |
| H$_2$O$_2$U | Ø40 | 0.145 | Al 250 μm | $^{235}$U | 99.999% | $6.18 \cdot 10^{17}$ | molecular plating |

The front-end electronics for the silicon detectors consisted of the MPR-16-LOG multichannel linear-logarithmic preamplifier, produced by Mesytec [24], followed by six ORTEC 474 timing filter amplifier modules [25]. The preamplifier has a linear behavior up to 10 MeV deposited energy, while above this energy its response becomes logarithmic. Its use allowed us to accommodate into a single range the low energy alphas and tritons as well as the highly energetic fission fragments and, especially, to minimize the effect of the so-called γ-flash, i.e. the prompt signal produced in the detector by relativistic particles and γ-rays from the spallation process. Such a signal was particularly large in this measurement, considering that the detectors were directly exposed to the beam. The data for each neutron bunch were collected by sampling the amplifiers' output signals by means of flash ADCs and recording the waveforms for 100 ms. In the following the six silicon detectors will be named as Li_f, Li_b, B_f, B_b, U_f, U_b, standing respectively for $^6$Li, $^{10}$B, $^{235}$U, forward and backward with respect to the beam direction.

## 2.1 Data analysis

The procedure for converting the time of flight into neutron kinetic energy made use of the high resolution time information taken from the U_f detector. A preliminary calibration of the flight path length was performed by means of a linear fit of the time-to-energy relation for forty prominent resonances between 2 and 35 eV neutron energy. The energy calibration was subsequently refined by minimizing the $\chi^2$ between the measured cross section and the ENDF-B/VIII evaluation, using a parametrization as in ref. [26], with a final path length L = 183.49 (0.02) m. A final check was performed by looking at the position of the aluminum dip, which resulted 5903.28 (1.05) eV to be compared with 5904.47 eV reported on all the major evaluated data libraries.

The stability of the silicon detectors was checked throughout the whole measurement, as their use in the high intensity neutron beam could have resulted in a degradation of their performances [27],[28]. The counting rates of the six detectors as a function of the neutron energy throughout the measurement (about one month of beam time) showed fluctuations of the order of 1% over the whole energy range, proving that the detectors remained stable and suffered no significant performance worsening.

Two-dimensional scatter plots of the deposited energy versus the incident neutron energy were built with the aim of selecting the reaction products, discriminating them from electronic noise and other background sources. Example of such plots are shown in Figure 3, Figure 4 and Figure 5, for Li_b, B_f and U_f respectively. In Figure 3 the two regions corresponding to the detection of alphas and tritons emitted in the $^6$Li(n,t) reaction are clearly distinguished. With increasing neutron energy, the number of counts decreases, as expected from the 1/v behavior of the cross section, while the deposited energy slightly decreases because of the kinematics, since the detector is positioned in the backward direction. In a similar fashion Figure 4 shows the data of the B_f detector, where two regions corresponding to the detection of α-particles and $^7$Li emitted in the n+$^{10}$B reaction can be clearly identified (in the upper part of the plot the higher energy α-particles from the (n,$\alpha_0$) reaction are also visible). Being this detector in the forward direction, the kinematics produces a slight increase in the deposited energy with increasing neutron energy.

In Figure 5 the same plot is shown for the U_f detector. The two regions corresponding to the detection of uncorrelated α-particles from the natural decay of $^{235}$U and fission fragments are clearly distinguished. The compressed range of the vertical axis is due to the logarithmic behavior of the preamplifier, which around and beyond 10 keV neutron energy starts to progressively lose gain because of the proximity to the very large γ-flash signal. Two bands close to each other are present in the fission region (upper part) of the plot, reflecting the mass (and energy) distribution of the fission fragments. In order to make sure that such a distortion was not producing a loss of fission fragments, we selected the centroid of the band in 7 neutron energy regions, and then adapted a polynomial function to these points. Such a function was used to straighten the plot that was then projected on the Y axis. In the resulting plot shown in Figure 6 (top) the two-bump structure of each bell-shaped curve clearly indicates the detection of fission fragments. We renormalized all the curves in order to have the same integral above the threshold (10500). The new plot, shown in Figure 6 (bottom), proves that the low energy tail is the same independently of the neutron energy, thus implying that the fraction of fission events lost below the threshold is the same at all energies. Consequently the efficiency of the selection cut for fission fragments was evaluated as 0.968 (0.0034).

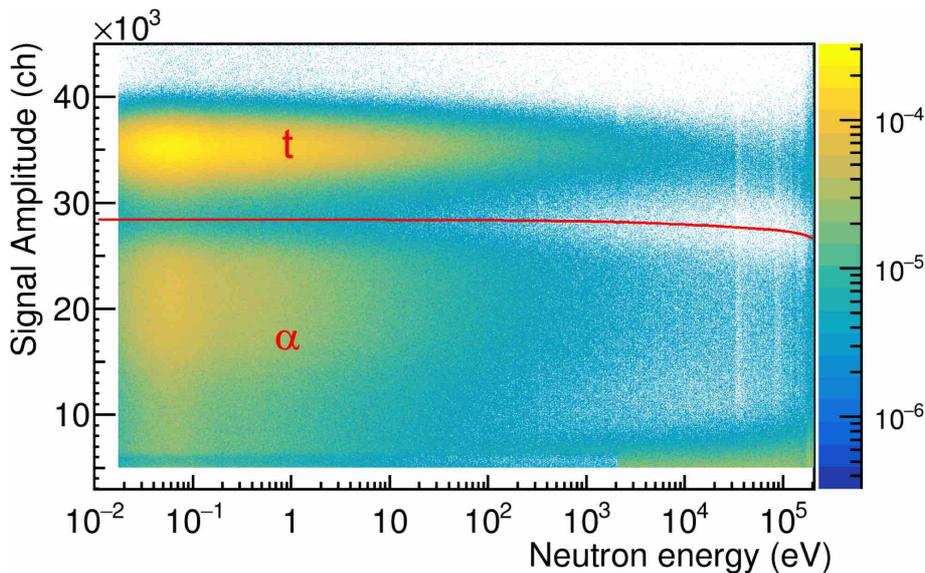

Figure 3 – (Color online) Scatter plot of deposited energy vs. neutron energy for the Li_b detector (backward emission from $^6$Li). The two regions corresponding to the detection of α-particles and tritons are clearly distinguished. The line represents the energy-dependent threshold applied to select only the triton region, used in the analysis (see text).

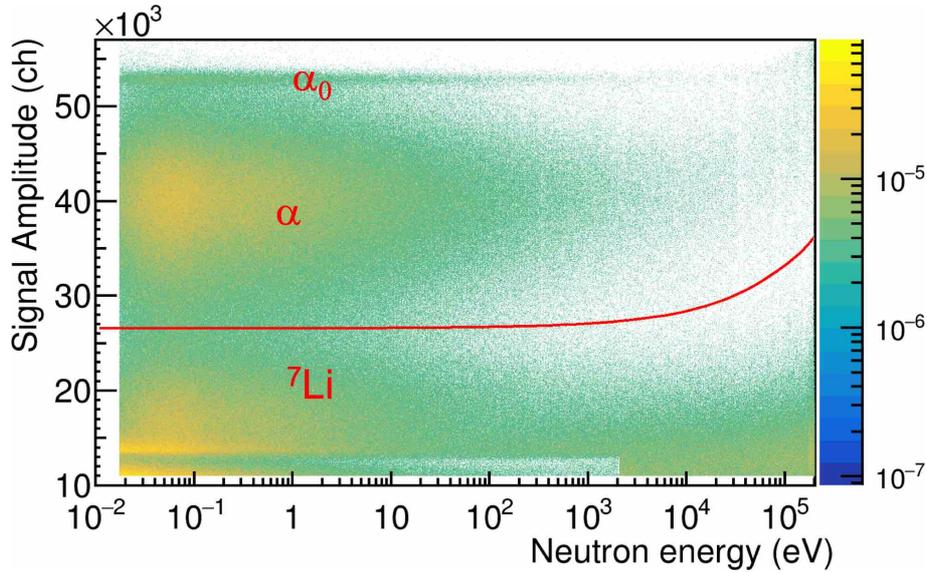

Figure 4 – (Color online) Scatter plot of deposited energy vs. neutron energy for the B_f detector (forward emission from $^{10}$B). The two regions corresponding to the detection of α-particles and $^7$Li are clearly distinguished (in the upper part also the higher energy α–particles from the (n,α$_0$) reaction are visible). The line is the energy-dependent threshold applied to select the α-particles used in the analysis (see text).

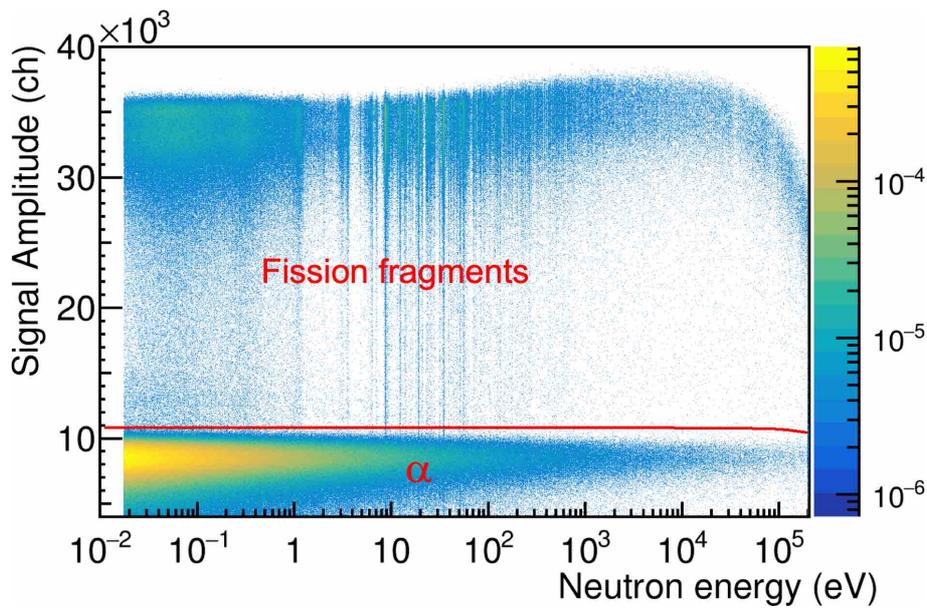

Figure 5 – (Color online) Scatter plot of deposited energy vs. neutron energy for the U_f detector (forward emission from $^{235}$U). The two regions corresponding to the detection of α-particles from the natural radioactivity of the sample, and fission fragments are clearly distinguished. The line shows the energy-dependent threshold applied to select the fission fragments for the analysis (see text). The logarithmic behavior of the preamplifier shows clearly in the compressed range of the vertical axis. The behavior of the amplitude at high neutron energies is related to the effect of the γ-flash.

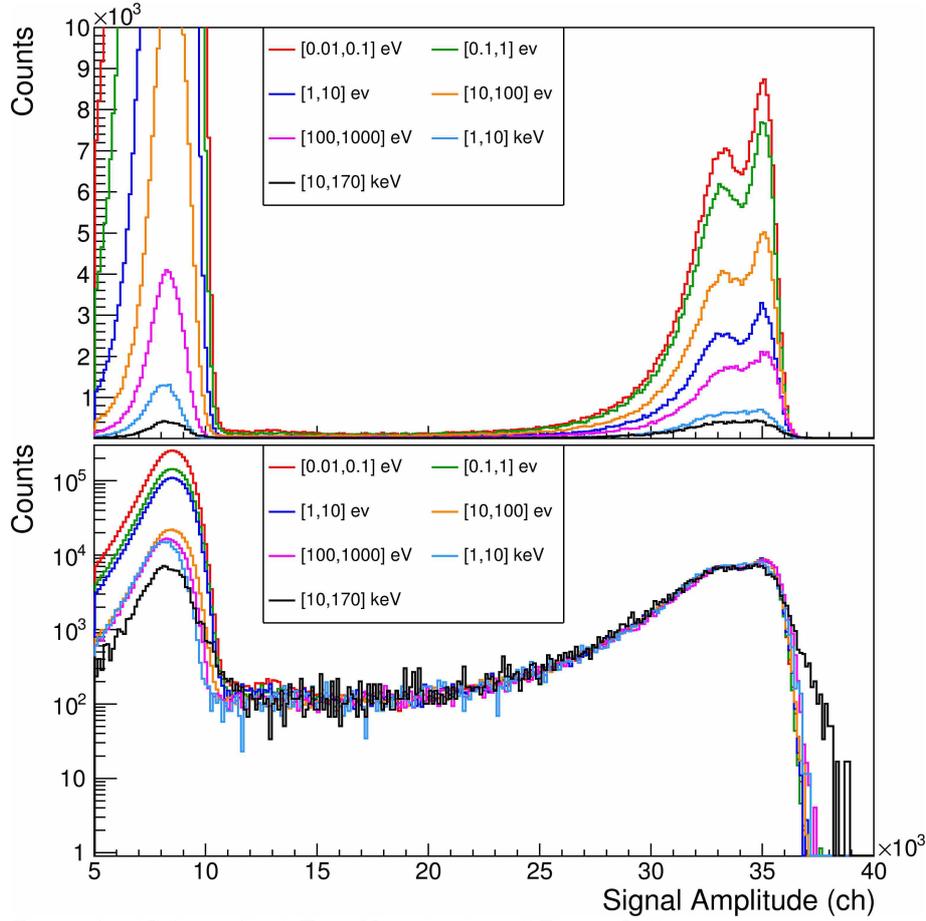

Figure 6 - (Color online) Top: Y-projection of Figure 5, for seven neutron energy intervals, after straightening the plot by means of a polynomial function. The two-bump structure of the bell-shaped curves proves that these are fission events. Bottom: The same plot, after renormalizing the curves to the same integral between 10500 (threshold) and 40000, shown in log-scale to prove that the fraction of fission fragments lost below the threshold is independent of the neutron energy (see the text).

The curves in the 2D plots for the reference reactions (Figure 3 and Figure 4) represent the conditions used to select the tritons and the α-particles for the $^{6}$Li(n,t) and $^{10}$B(n,α) reactions, respectively. These curves were obtained by analyzing the 1D amplitude spectra for different neutron energies. The optimal identification threshold, determined as the minimum in the valley between the reaction products (Figure 7), was fitted as a function of energy with a polynomial function. The same could not be done for the Li_f detector since, due to kinematical reasons and a slightly worse resolution, at increasing neutron energy the two structures from tritons and α-particles were not clearly separated. In this case the threshold was safely chosen at -1 standard deviation from the maximum of the triton peak. For the U_f and U_b detectors, the threshold for each neutron energy was chosen just above the α-particle peak, in a position where the counts drop by two to three orders of magnitude (see also Figure 6). A polynomial fit similar to the previous ones provided the analytical energy-dependent threshold curve (the one for the U_f detector is shown in Figure 5).

The residual background surviving the amplitude cuts was measured by means of dummy samples, i.e. only made of the backing, and it was found very small as compared to the reaction count rate, therefore it was neglected in the analysis. No background could be expected from C, O and F present in the samples, as their (quite small) cross sections for the production of charged particles have thresholds of the order of several MeV.

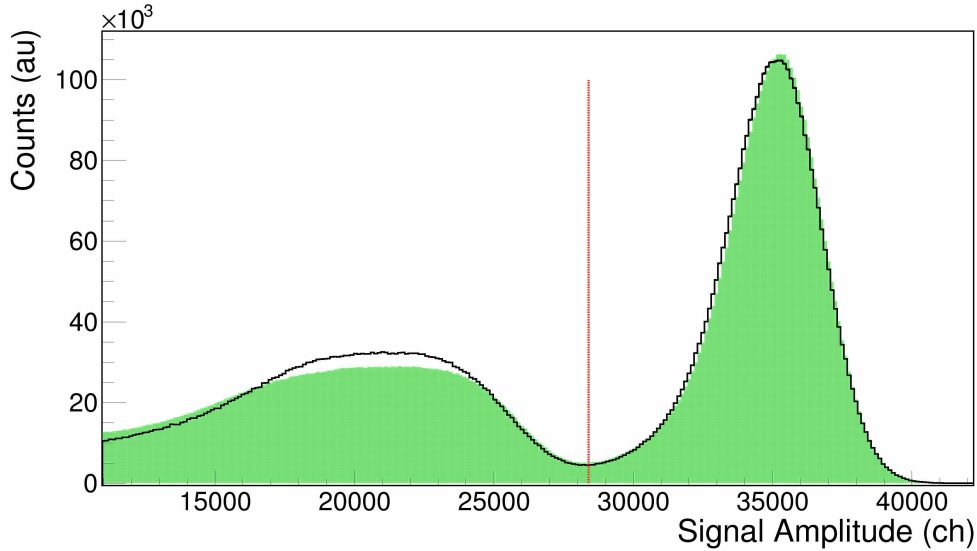

Figure 7 - (Color online) Deposited energy spectrum for the Li_b detector at thermal neutron energy. The two structures correspond to the detection of tritons and alphas. Filled area: experimental data. Line: simulation result. Vertical line: identification threshold used in the analysis.

### 2.2 Neutron beam fraction and detection efficiency

The measurement aimed at determining the $^{235}$U(n,f) cross section relative to the standard ones of $^{6}$Li(n,t) and $^{10}$B(n,α) reference reactions. This goal can be achieved by means of the so-called "ratio method", according to the following equation:

$$\sigma_{^{235}U} = \frac{C_{^{235}U}\, f_{ref}\, \rho_{ref}\, \epsilon_{ref}}{C_{ref}\, f_{^{235}U}\, \rho_{^{235}U}\, \epsilon_{^{235}U}} \sigma_{ref} \qquad (1)$$

Here, $C_X$ is the number of counts for a given sample X, $\rho_X$ is the areal density for that sample, $f_X$ is the neutron beam fraction intercepting it and $\epsilon_X$ the detection efficiency for the corresponding reaction products. All these quantities, apart from $\rho_X$, are a function of neutron energy. The main advantage of the ratio method is that it allows one to almost completely neglect the neutron fluence and its energy dependence, because the neutron beam incident on the $^{235}$U and reference samples is practically the same. Small differences in the neutron beam impinging on each sample, related to the geometrical shape of the sample and to the neutron beam absorption along the setup (i.e. in the various samples and detectors) are taken into account by the correction factor $f_X$. This was estimated by Monte Carlo simulations of the neutron interaction with all the layers of different materials encountered along its trajectory. To this purpose the full geometry of the apparatus was implemented in detail in the GEANT4 Monte Carlo code [29]. $1.25 \cdot 10^7$ incident neutrons per decade were randomly generated in the 10 meV to 1 MeV energy range, each decade divided into 1000 bins, with a uniform probability inside each bin. The geometrical distribution of these neutrons in the transverse plane was gaussian with a standard deviation of 7 mm, reflecting the known spatial profile of the neutron beam. The propagation of the simulated beam through the experimental apparatus allowed us to evaluate the effective fraction of neutrons impinging on each sample. The results for the different samples are shown in Figure 8. Between 1 eV and 30 keV the neutron beam fraction is rather flat for all samples, whereas in the thermal region, and above 30 keV where the capture resonances in the silicon and in the aluminum backings set in, the absorption may reach several percent, and is particularly large for the last sample and for the main Al and Si resonances.

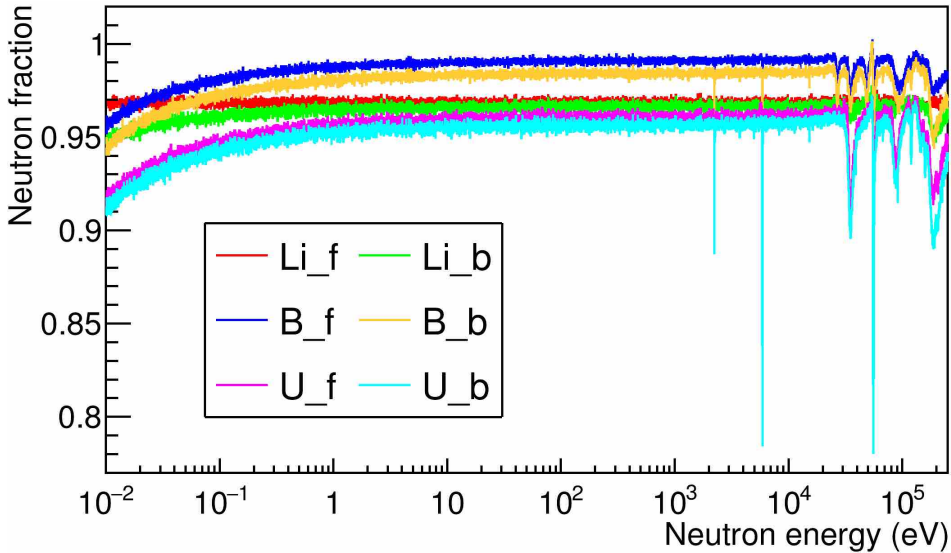

Figure 8 – (Color online) Simulation of the fraction of the neutron beam impinging on each sample. Between 1 eV and 30 keV neutron energy, the neutron absorption is rather flat. In the thermal region, and above 30 keV where the capture resonances in the silicon and in the aluminum of the sample backing set in, the absorption reaches several percent.

The detection efficiency $\varepsilon_X$ in Eq. (1) needs also to be determined with high accuracy for all measured reactions, taking into account the geometrical efficiency of the silicon detector as well as the detection threshold and other conditions used in the analysis. To this aim, the same Monte Carlo simulations previously described were also employed. The only difference is that instead of starting from a neutron beam impinging on the sample, which would have been time consuming due to the tiny reaction probabilities, the GEANT4 simulations were performed by generating directly the reaction products, uniformly emitted from the reference samples according to the transverse beam profile. In the simulations, the angular distribution of the particles emitted from the samples was generated according to ENDF-B/VIII data ([10],[30]), for twelve neutron energies from thermal to 170 keV.

In each of the twelve simulations, the neutron energy was also considered to account for the kinematic boost of the reaction products. For each of the $^6$Li and $^{10}$B samples, and for each neutron energy, $10^5$ reactions were simulated, with the reaction products from each sample transported until they hit the detectors or exited the experimental setup. The energy deposited in the silicon detectors was recorded and a suitable resolution was applied so as to reproduce the measured amplitude spectra, as can be seen for instance in Figure 7 for the Li_b detector. As for the $^{235}$U sample, the neutron energy is negligible compared with the kinetic energy of the fission fragments and the angular distribution is isotropic. As mentioned in Section 2.1 and shown in Figure 6, the detection efficiency for fission fragments does not depend on the neutron energy, therefore it was decided to leave $\varepsilon_U$ as an unknown constant and to normalize the final cross section to a standard value (see also Section 3.2).

The detection efficiency depends on the threshold used in the analysis, i.e. the one shown by the curves in Figure 3 to Figure 5. In order to use in the simulations a threshold consistent with the one used on the data, the amplitude spectra for the detectors Li_f, Li_b, B_f and B_b were calibrated in energy by fitting them to the simulated ones, thus allowing us to calculate the energy-dependent thresholds in energy units. Finally, a polynomial fit of the efficiency as a function of the neutron energy for the twelve simulated points provided a reliable analytical form of the detection efficiency, as shown in Figure 9 for Li_f, Li_b, and in Figure 10 for B_f, B_b. The reduced efficiency of Li_f in Figure 9 is a consequence of the high threshold previously mentioned, which rejects a fraction of the tritons in the data analysis. Above about 1 keV neutron energy the efficiency in the forward direction increases, with a corresponding decrease in the backward direction, as expected from the kinematic boost. Above 10 keV the angular distribution of tritons from the $^6$Li(n,t) reaction becomes forward peaked, due to the p-wave resonance at 235 keV, and this affects the shape of the efficiency of the Li_f and Li_b detectors (Figure 9). An overall check of this procedure, done by comparing the plots in Figure 9 and Figure 10 with the expected distributions from ENDF-B/VIII for two single angles (0° and 180°) showed a very similar behavior.

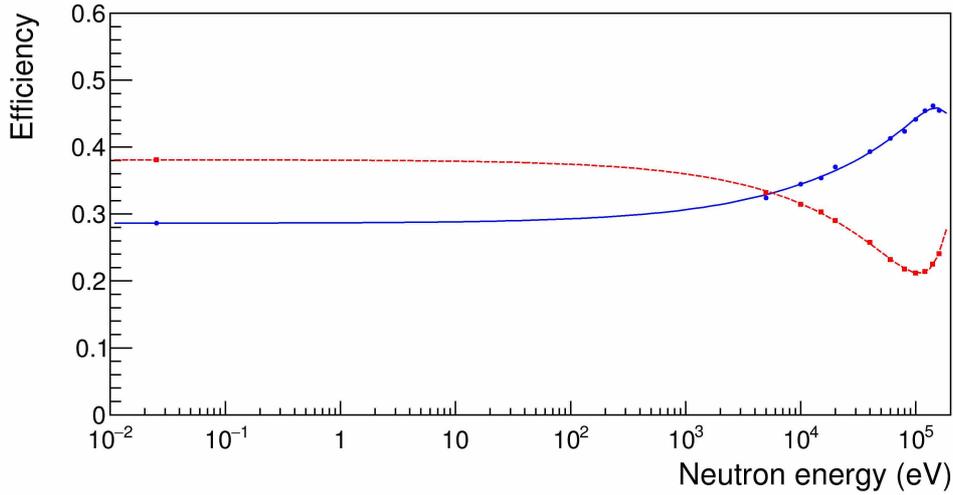

Figure 9 - Detection efficiency as a function of the neutron energy for the Li_f (solid line) and Li_b (dotted line) detectors. The reduced efficiency of Li_f is due to the higher threshold adopted, which only considers part of the tritons for the data analysis (see text).

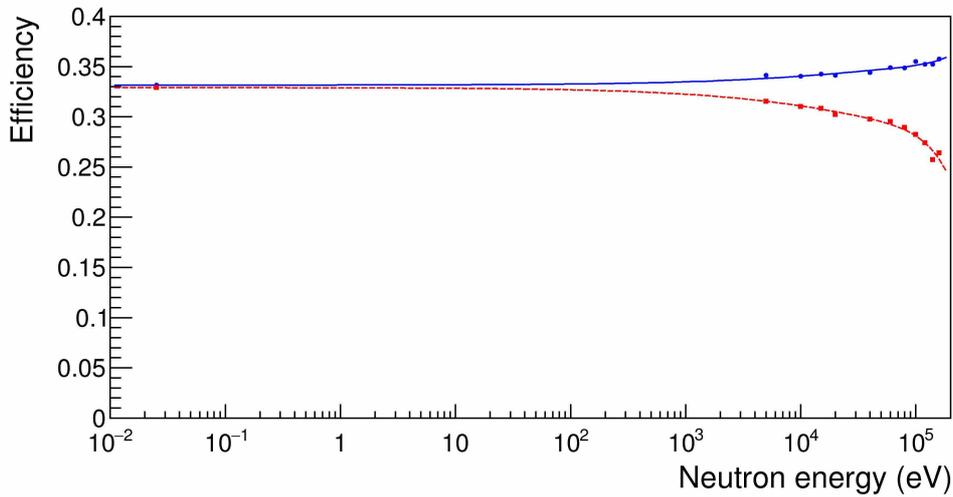

Figure 10 - (Color online) Detection efficiency as a function of the neutron energy for the B_f (solid line) and B_b (dotted line) detectors.

### 2.3 Systematic uncertainties

Two main possible sources of systematic uncertainty were investigated, namely the choice of the event selection cuts and the alignment of the setup with respect to the beam. The former influences the number of counts (see Eq. 1) for the individual reactions, as well as the calculated detection efficiency, whereas the latter could bear some impact on the fraction of the neutron beam impinging on the various samples. The data analysis and Monte Carlo efficiency calculations were performed with six different cuts obtained by shifting the nominal energy-dependent threshold by ±1%, ±2%, ±3% (which lead to relevant variations in terms of counts). Nonetheless, the variation of the reaction yields corresponding to the ±3% shift of the amplitude cuts (worst case scenario) is ±0.3% for the $^6$Li data and ≤1% for $^{10}$B data. No appreciable effect was observed for the $^{235}$U data.

The experimental apparatus was initially aligned to the nominal beam position. However, the real position of the neutron beam was then checked throughout the measurement by means of a photographic emulsion (Gafchromic foil) placed in front of the vacuum chamber hosting the setup. The effective beam center position was found to be displaced by 10.0 (0.7) mm and 5.0 (0.7) mm, respectively in the horizontal and vertical direction, with respect to the detector center. This was adopted as the reference beam center for the efficiency simulations and data analysis. To estimate the uncertainty related to the detector alignment, the data analysis and Monte Carlo efficiency calculations were performed again by shifting the reference beam center up to ±2 mm (about 3 standard deviations) in X and Y. We verified that such a shift leads to a slight change of efficiency which is roughly linear with the displacement, resulting ≤1% for a shift of one standard deviation. These possible sources of systematic uncertainty, combined in quadrature, point to a reasonable (and conservative) estimate for the systematic uncertainty of about ±1.5%.

## 3 Results
### 3.1 The $^6$Li(n,t) to $^{10}$B(n,α) cross section ratio

A validation of the experimental technique and analysis method can be obtained from the results of the $^6$Li(n,t) and $^{10}$B(n,α) reactions. In particular, although the determination of the $^{235}$U(n,f) cross section does not require the knowledge of the neutron flux, it is useful in this context to reconstruct it from the data relative to the two reference reactions (the term "neutron flux" refers here for simplicity to the energy distribution of the total number of neutrons in a bunch, a quantity that should be more appropriately called "instantaneous intensity"). In this work, the neutron flux was reconstructed independently with each of the four Li_f, Li_b, B_f and B_b targets and detectors, according to the following expressions:

$$\Phi_{Li\_f} = \frac{C_{Li\_f}}{f_{Li\_f} \cdot (1-e^{-\rho_{Li\_f} \cdot \sigma_{Li\_f}}) \cdot \varepsilon_{Li\_f}}; \quad \Phi_{Li\_b} = \frac{C_{Li\_b}}{f_{Li\_b} \cdot (1-e^{-\rho_{Li\_b} \cdot \sigma_{Li\_b}}) \cdot \varepsilon_{Li\_b}} \quad (2)$$

$$\Phi_{B\_f} = \frac{C_{B\_f}}{f_{B\_f} \cdot \rho_{B\_f} \cdot \sigma_{B\_f} \cdot \varepsilon_{B\_f}}; \quad \Phi_{B\_b} = \frac{C_{B\_b}}{f_{B\_b} \cdot \rho_{B\_b} \cdot \sigma_{B\_b} \cdot \varepsilon_{B\_b}} \quad (3)$$

Here C represents the number of counts normalized to the nominal proton bunch of 7x10$^{12}$ protons, and σ the standard cross sections for the two reference reactions (taken from the IAEA reference file [31]), with all the other factors defined as in Eq. (1). Except for the areal density, all quantities in the expressions are a function of neutron energy. Since the two Li samples are rather thick, the self-absorption of the neutron beam in the sample is taken into account in Eq. (2) (by means of a simplified expression), while on the contrary, for both the $^{10}$B and $^{235}$U samples, such a correction is very small, of the order of 10$^{-4}$, and can therefore be neglected. Ideally, the four expressions above should give the same flux. However, while the shape (i.e. the neutron energy dependence) is similar for the four detectors, a few percent difference in the absolute value is observed. A comparison of the integral in the 1÷10 eV range leads to a ratio $\Phi_{Li\_f}/\Phi_{Li\_b}$=0.967 and $\Phi_{B\_f}/\Phi_{B\_b}$=1.109, hinting at a difference of about 3% between the areal densities of the two $^6$Li samples and of about 10% between the $^{10}$B ones.

To minimize the effect of the uncertainty on the areal density, the four different results on the neutron flux were all normalized to each other in the 1÷10 eV range. A weighted average was then calculated for the two $^6$Li and the two $^{10}$B samples, so as to obtain a unique value for each of the two reference reactions (referred to hereafter as $\Phi_{Li}$ and $\Phi_B$). The corresponding results are shown in Figure 11, where an almost perfect agreement can be seen.

For a more quantitative comparison between the two different results, the following three statistical indicators were constructed, as a function of the neutron energy:
- The ratio between the flux extracted on the basis of the two reference reactions, which provides a numerical indication of the mutual deviation in relative units.
- The normalized deviation Σ between the two distributions, in standard deviation units, rebinned in groups of 50 points so to increase its statistical significance. This quantity provides a numerical indication of statistically relevant systematic deviations between two distributions.
- The reduced χ$^2$ between the two distributions, calculated in groups of 50 points, which provides information on their mutual agreement (typically between data and a reference model). χ$^2$ is quite useful as an indication of the shape (dis)agreement between the two distributions mentioned above, where we took as model the $^6$Li data because of the better statistics.

The following expressions were used for the two statistical indicators mentioned above:

$$\chi_k^2 = \frac{1}{50}\sum_{i=1}^{50}\frac{(y_i-Y_i)^2}{\sigma_{y_i}^2} \qquad \Sigma_k = \frac{\sum_{i=1}^{50}(y_i-Y_i)}{\sqrt{\sum_{i=1}^{50}\sigma_{y_i}^2}} \quad (4)$$

Eqs. (4) indicate how χ$^2$ and Σ are calculated in a wider bin $k$ by means of 50 consecutive smaller bins indexed by $i$. Here $y$ stands for the data values with uncertainty $\sigma_y$, Y for the model values. These indicators are quite useful to identify possible statistically significant deviations between two distributions, and are reported in the three panels of Figure 11. In particular, the ratio $\Phi_B/\Phi_{Li}$ is shown as a function of energy, along with the normalized deviation Σ in standard deviation units and the reduced χ$^2$ of $\Phi_B$ with respect to $\Phi_{Li}$. The ratio $\Phi_B/\Phi_{Li}$ is close to 1 within ≈1% up to 1 keV, while fluctuations of several percent are present above a few keV, due to the higher statistical uncertainty in that region for $\Phi_B$, in consequence of the small mass of the $^{10}$B samples. As expected, the χ$^2$ is uniformly distributed around 1, while the normalized deviation Σ keeps symmetrical around ±1÷2 standard deviations, proving that the two independent reconstructions of the incident neutron flux are consistent with each other. It is worth noticing that in the neutron energy region below 100 meV the quantity Σ points to a statistically significant difference between the B and Li results. However, this difference is below 1%, likely due to the uncertainty in the neutron beam fraction, and still below the systematic uncertainty.

While the results discussed above show the correctness of the data for the reference reactions, a more direct evidence of the reliability of the $^{10}$B and $^6$Li data can be obtained by constructing the ratio of the respective count rates (corrected by the neutron beam fraction and efficiency previously described). Figure 12 shows such a ratio, compared with the ratio between the $^{10}$B(n,α) and $^6$Li(n,t) standard cross sections (from the evaluated data files [31]). In this case as well, the relatively large uncertainty related to the sample thickness has been eliminated by normalizing the measured ratio to the evaluated one, in the 1÷10 eV neutron energy range.

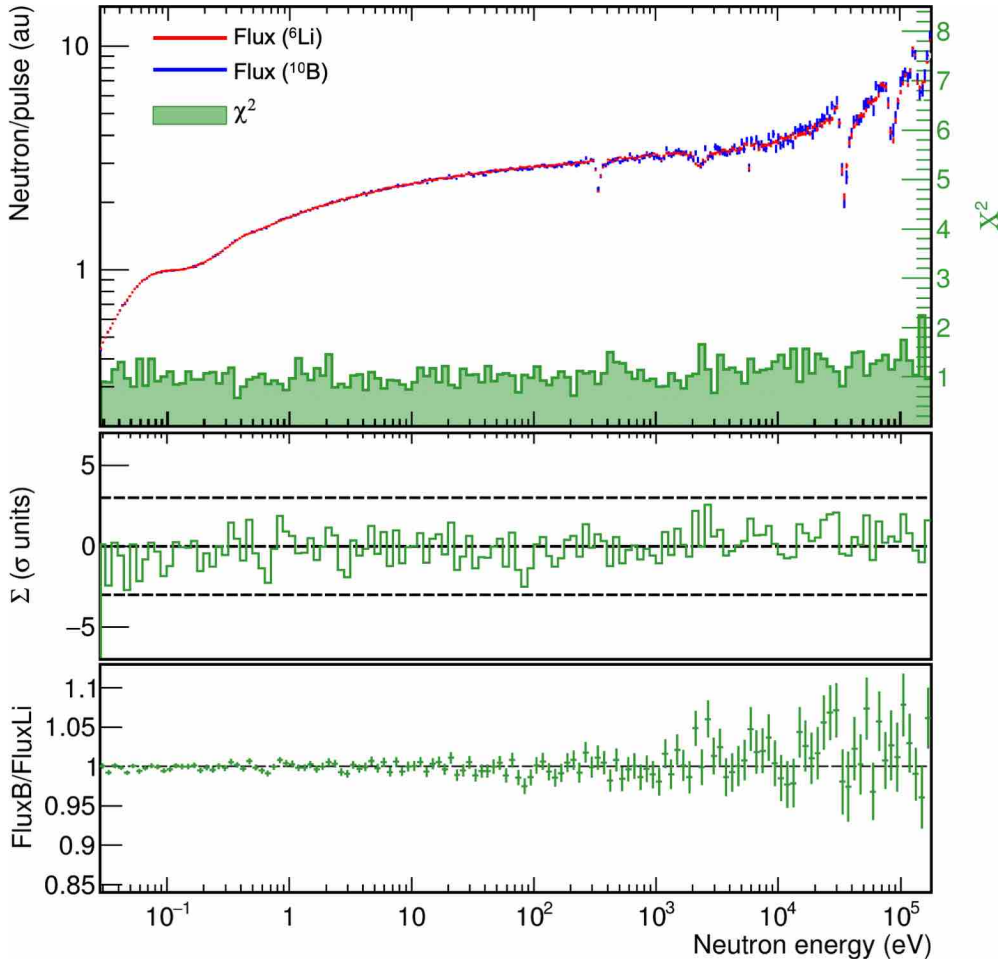

Figure 11 – (Color online) The neutron flux as evaluated in the present work by using data from the $^6$Li and the $^{10}$B samples. See the text for the meaning of the other statistical indicators.

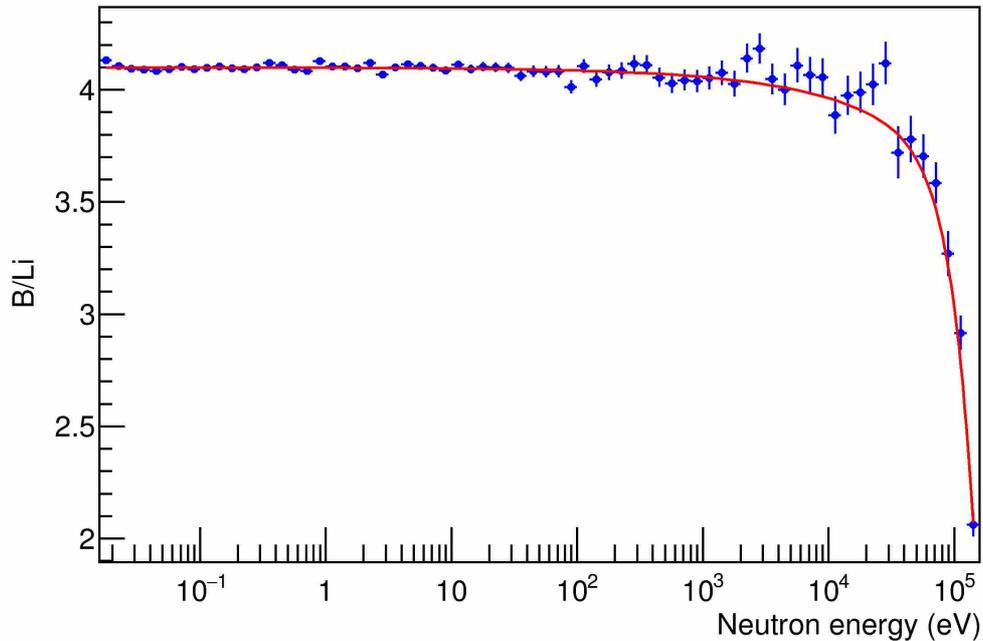

Figure 12 – (Color online) Ratio of the corrected $^{10}$B to $^{6}$Li count rates (dots), along with the corresponding ratio between the standard cross sections from the evaluated data files [31] (line).

### 3.2 The $^{235}$U(n,f) cross section

As in the case of the reference reactions, to determine the $^{235}$U(n,f) cross section the weighted average of the count rate in the backward and forward detector $C_{U\_b}$ and $C_{U\_f}$, respectively, was computed. In this case, the two distributions are virtually identical (i.e. no statistically significant difference between the two data sets is observed, as expected in consideration of the isotropic emission of the fission fragments). Two different $^{235}$U(n,f) cross sections were computed according to Eq.(1), relative to the $^{6}$Li(n,t) and to the $^{10}$B(n,α) reactions using the corresponding weighted average of the forward and backward distributions. The individual cross sections shown throughout this work have been calculated at 1000 bin/decade. For the sake of simplicity all results shown in the following refer to a single $^{235}$U(n,f) cross section obtained from a single reference flux, computed as weighted average between $\Phi_B$ and $\Phi_{Li}$.

In order to rule out the unknown detection efficiency $\varepsilon_U$ the three versions of the cross section were normalized to a standard value of the $^{235}$U(n,f) cross section from libraries, thus also removing the uncertainty due to the sample thickness. The normalization constants can be chosen in order to match the cross section value at the thermal point, which is a standard. However, as this would be a single point, such a normalization constant would be affected by a large statistical uncertainty reflecting into a large systematic uncertainty in the whole cross section. Conversely, a normalization to a suitable integral, having a much smaller statistical uncertainty, produces a much smaller systematic uncertainty. Therefore the normalization to the integrated cross section in the energy interval 7.8÷11.0 eV was preferred, as recommended by IAEA [8]. It should be noticed, however, that the two normalization methods are consistent with each other, within the statistical uncertainty. This can be clearly seen in Figure 13, where the ratio between the thermal cross section and the integrated one in the 7.8÷11.0 eV range is reported for several evaluations and for the data of this work. The present results agree with the recommended IAEA, ENDF/B-VIII and JEFF3.3 values, and are in all cases within 1%.

The following remarks summarize the calculations described above and the corresponding results:

- The independent normalization of the count rate of the reference reactions in the 1-10 eV region, eliminates completely the systematic uncertainty related to the areal density of the used reference samples, or their possible inhomogeneity. Furthermore, normalizing in this energy range, rather than at thermal energy, further minimizes the overall uncertainty, considering the larger uncertainty in the thermal neutron energy region previously seen (Figure 11).
- An additional systematic uncertainty is related to the angular distributions of the products in the $^{6}$Li(n,t) and $^{10}$B(n,α) reactions as a function of neutron energy. Indeed, even though reasonable evaluations are available [10], the true angular distributions are not exactly known. This has an impact on the efficiency calculations. Due to possible differences between ENDF-B/VIII and the real angular distributions, slight residual differences between results from the forward and backward detectors could still be present. These possible differences were however further smeared out by using the weighted average between forward and backward data, thus obtaining two distinct reference data sets (Lithium and Boron).

- These two data sets, after checking their mutual statistical consistency (Figure 11), were further combined into a weighted average in order to be used as a final reference for the $^{235}$U(n,f) cross section in all the following plots.
- A further normalization of the ratio between the $^{235}$U(n,f) data and the combined ones for the reference reactions was performed relative to the integral in the 7.8÷11.0 eV range, thus eliminating the uncertainty on the areal density (and inhomogeneities) of the $^{235}$U sample and on the detection efficiency $\varepsilon_U$.

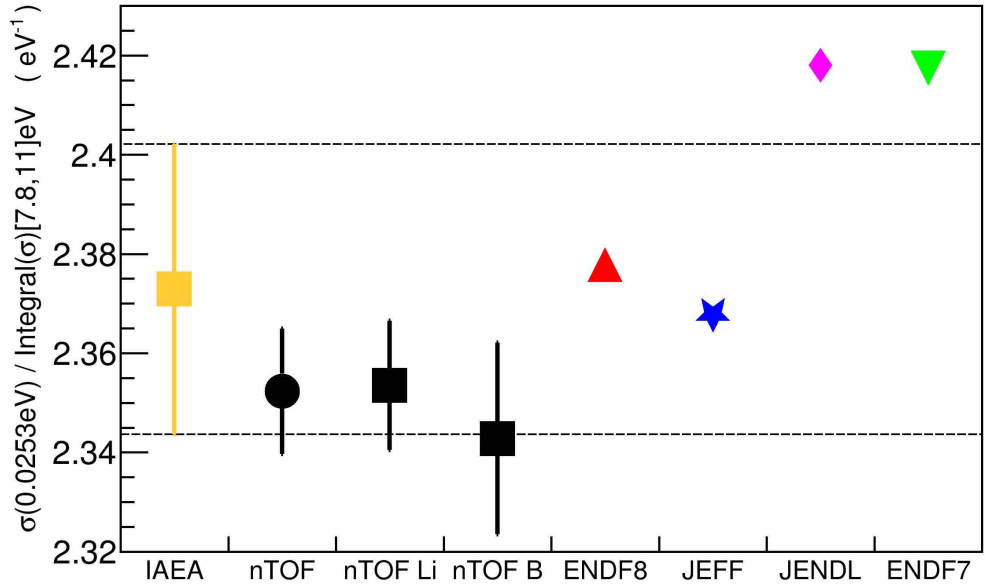

Figure 13 – (Color online) Ratio between the $^{235}$U(n,f) cross section at the thermal point and the integrated cross setion in the 7.8÷11.0 eV neutron energy interval for several evaluations and the data of this work, separately with respect to Li and B, and their weighted average.

Based on the above considerations, it can be estimated that the $^{235}$U(n,f) cross section (or cross section ratio relative to the individual reference reactions) are affected by an overall systematic uncertainty of about 1.5 % in the whole energy region. The $^{235}$U(n,f) cross section $\sigma_{235U}$, in the full energy range explored (0.02 eV ÷ 170 keV), is shown in the top panel of Figure 14. The lower part of the panel shows the reduced $\chi^2$ with respect to the ENDF-B/VIII and JEFF3.3 evaluations, the middle panel contains the normalized deviation Σ between the current results and the libraries, as shown and discussed in better detail in the following section, whereas the lower panel shows the ratios of the current data to the two libraries. The same plots for the ENDF-B/VII and JENDL libraries are shown in Figure 15.

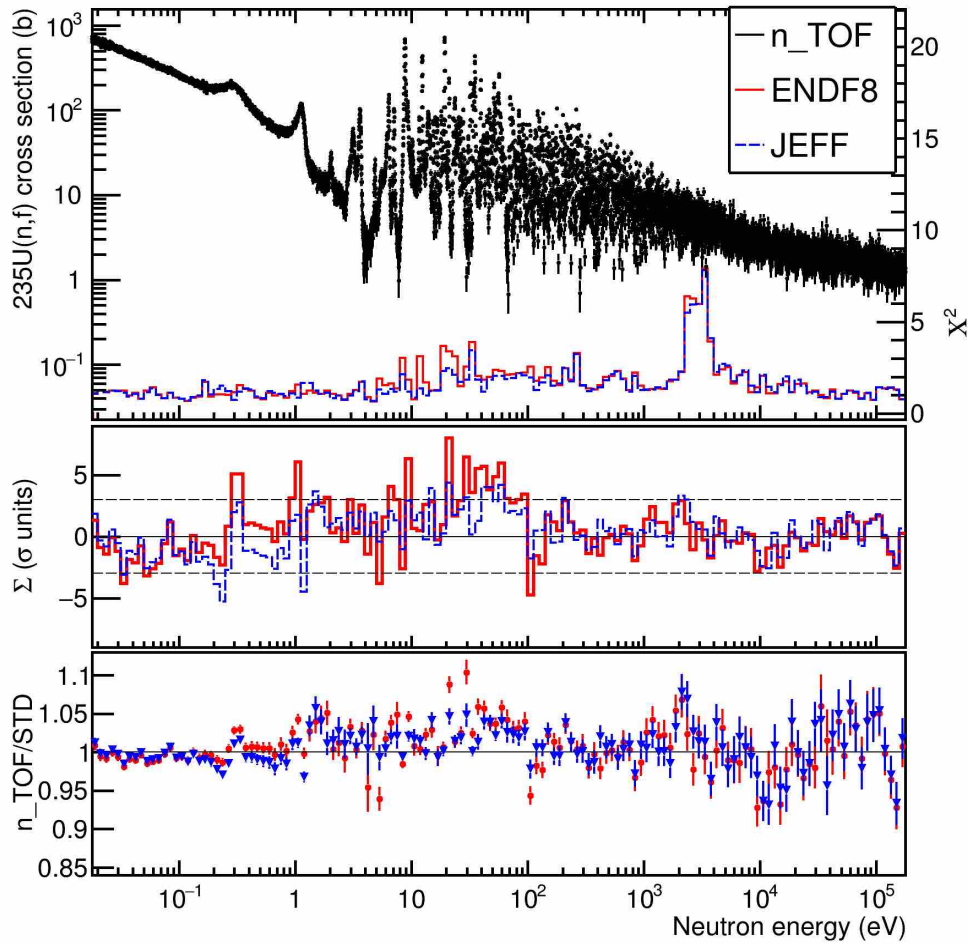

Figure 14 – (Color online) Top panel: the final measured $^{235}$U(n,f) cross section of this work, obtained from the ratio method relative to the weighted average of the $^{6}$Li(n,t) and $^{10}$B(n,α) data; in the lower part the reduced $\chi^2$ with respect to the ENDF-B/VIII and the JEFF3.3 evaluations is shown. Middle panel: the normalized deviation Σ between the current data and the two libraries; the dashed lines indicate the ±3σ level. Bottom panel: the ratio of the current data to the two libraries.

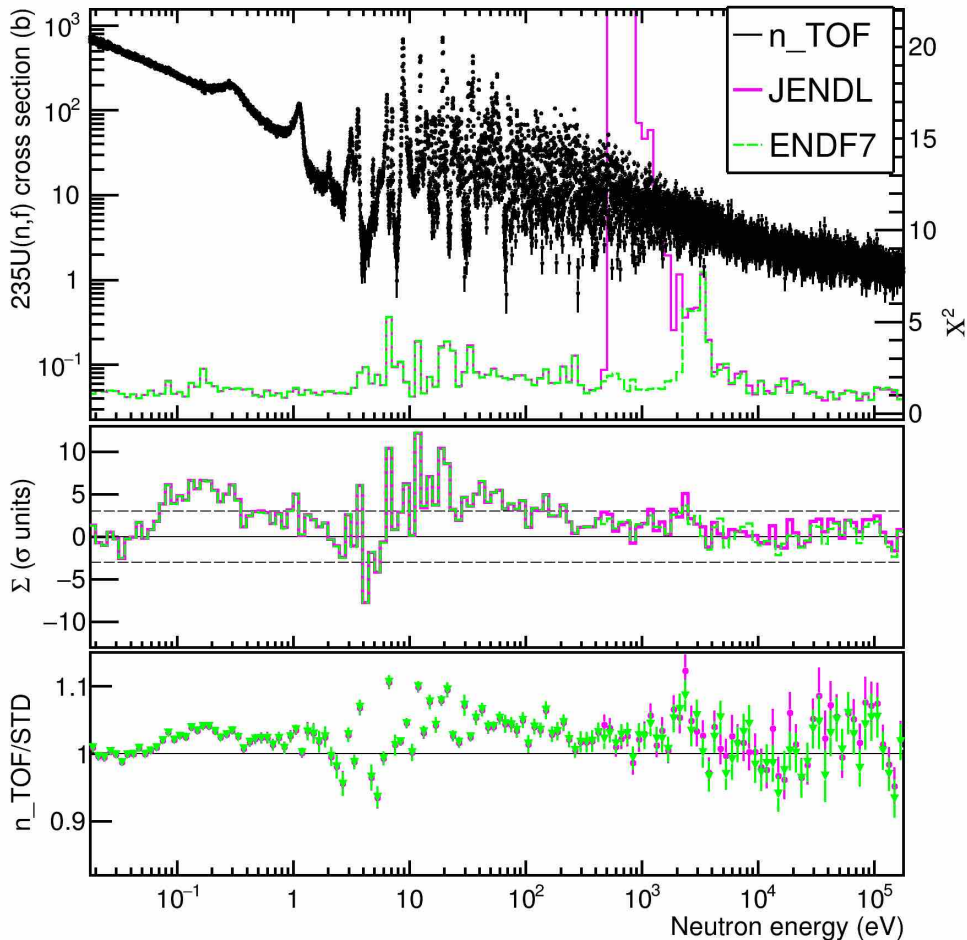

Figure 15 – (Color online) Top panel: the final measured $^{235}$U(n,f) cross section of this work, obtained from the ratio method relative to the weighted average of the $^6$Li(n,t) and $^{10}$B(n,α) data; in the lower part the reduced $\chi^2$ with respect to the ENDF-B/VII and the JENDL evaluations is shown. Middle panel: the normalized deviation Σ between the current data and the two libraries; the dashed lines indicate the ±3σ level. Bottom panel: the ratio of the current data to the two libraries.

## 4 Discussion

Even though the initial purpose of the present experiment and data analysis was to investigate in better detail the cross section in the neutron energy range around 10÷30 keV, the combination of the convenient features of the n_TOF neutron beam with the high performance of the newly developed experimental apparatus and technique has led to high accuracy, high-resolution data in a much wider range, from thermal neutron energy to 170 keV. A few considerations can be made on the basis of the present results.

First of all, the middle and bottom panels of Figure 15 indicate that the evaluations ENDF-B/VII and JENDL fail to reproduce the current data in several regions: both at low energy (<1 eV) and in the resolved resonance region up to 200 eV the quantity Σ indicates a statistically significant deviation which is also visible in the bottom panel as a shift of several percent. This is not the case with the more recent evaluations (ENDF-B/VIII and JEFF3.3) of Figure 14, where |Σ|≤ 3 almost everywhere, especially for JEFF3.3. In particular, up to 1 eV neutron energy the present data agree with the recent evaluations within 1%, whereas differences of several percent (up to 10% in some regions) are observed in the resolved resonance region, up to 100 eV. Such differences are most likely related to local mismatches in the amplitude or shape of some resonances, which lead to a corresponding sharp fluctuation of Σ (mismatched amplitude) or to a peak in $\chi^2$ (mismatched shape), thus probably indicating a lower accuracy of the evaluation in that region. From 100 eV to 10 keV a reasonable agreement, of the order of 2%, is again observed.

The large $\chi^2$ between data and evaluations observed in Figure 14 in the 2÷4 keV region is simply related to the presence, in the n_TOF data, of resonance-like structures not reported in the evaluations (we remark here that the boundary between resolved and unresolved resonance regions in the evaluations is somewhat arbitrary). For the same reason a very large peak in $\chi^2$ is visible even at lower energy in Figure 15 for the JENDL evaluation, starting at 500 eV that is where the resolved resonance region stops in such library. Finally, in the region from 30 to 100 keV differences of several percent are observed between the present data and all evaluations, although with a relatively low statistical significance because of the larger statistical error. In the following we will focus on the region above 2.2 keV and in particular up to 30 keV, leaving to a forthcoming paper a more detailed analysis of the resolved resonance region.

A zoom of Figure 14 in the neutron energy range from 1 to 10 keV is shown in Figure 16. As previously mentioned, above 2.25 keV the evaluations do not report any structures but rather an average behavior. This is likely due to the lack of experimental data [32] or perhaps to the fact that in the evaluation process these structures might have been attributed to statistical fluctuations in the data sets adopted as reference. On the contrary, the present results unambiguously show that several structures do exist, as proved by the behavior of the statistical indicators, with the $\chi^2$ rising up to 8 between 2.25 and 4 keV to indicate a definite shape mismatch (the $\chi^2$ behavior is the same even if calculating it with a different binning). However, the $\Sigma$ behaviour, with no relevant deviation, signals that on average the data trend correctly follows the latest evaluations. These indicators clearly show that the observed structures are significant and not due to statistical fluctuations.

In order to prove this, similarly to the procedure exploited in ref.[19], the level sequence of compound states distribution in $^{235}$U+n was simulated just above the neutron separation energy S(n) = 6.544 MeV, assuming a GOE surmise [33] and adopting an average level spacing $D_0$ = 0.54 eV. From the generated sets of neutron resonances the neutron induced fission cross section for $^{235}$U was simulated. It can be shown that the various components of the energy resolution at EAR1 can be converted into an effective temperature, assuming that the shape of the resolution is gaussian, and that the effective temperature corresponding to $\approx$3 keV neutrons is about 983°K. The plot of Figure 17 (top) clearly shows the resonance grouping resulting from this simulation, which is qualitatively compatible with the observed structures in the measured data (Figure 17 bottom). Above 4 keV the $\chi^2$ decreases, because of the increasing level density and worsening of the resolution, and therefore no significant statement can be made about possible structures in the cross section.

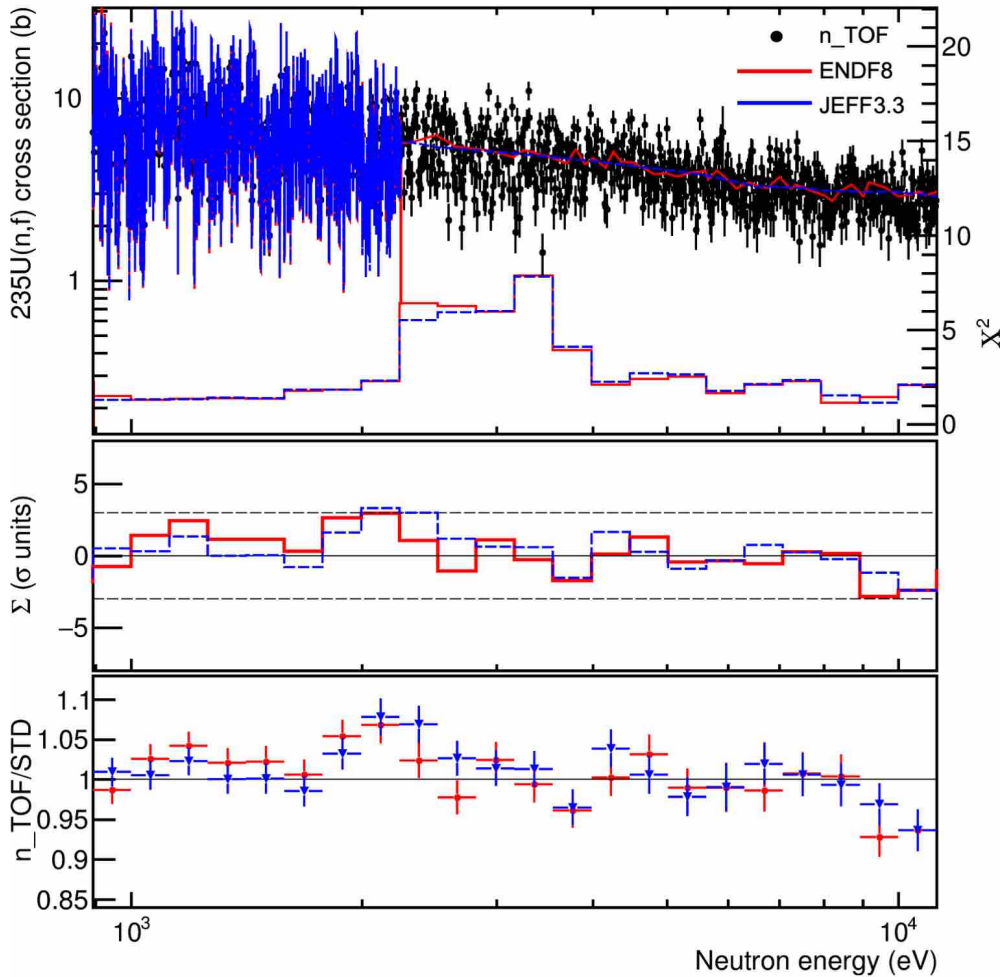

Figure 16 - (Color online) Top panel: $^{235}$U(n,f) cross section of this work (with $\chi^2$, $\Sigma$ and ratio), in the 1÷10 keV neutron energy range, along with the corresponding data from the ENDF-B/VIII and the JEFF3.3 evaluations.

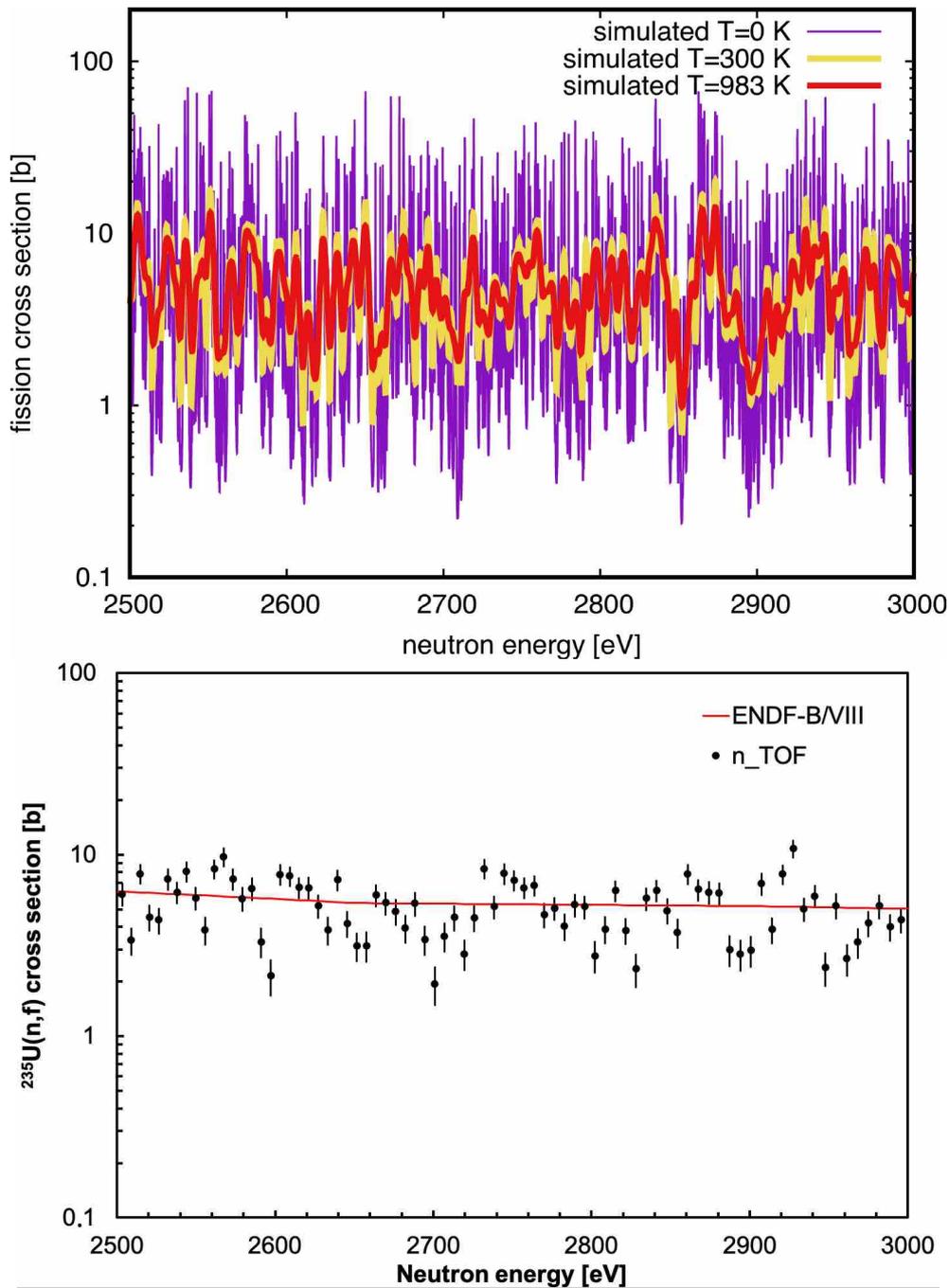

Figure 17 - (Color online) Top: simulation of the $^{235}$U(n,f) cross section for three different effective temperatures, corresponding to different FWHM resonance resolution. Given the realistic effective temperature of 983°K expected at ≈3 keV, a resonance grouping shows up clearly (see the text). Bottom: n_TOF and ENDF-B/VIII cross section data in the same energy range.

Between 8 and 100 keV the present data show systematic deviations from the evaluations, as can be clearly observed in Figure 18. The trend is not unique: below 30 keV the present data are lower than the evaluations by up to 8%, while between 30 and 100 keV the measured n_TOF cross section is slightly higher. The observed discrepancy in the 10-30 keV region seems to confirm the previous indication of a shortcoming of major evaluated data libraries in that energy region [1].

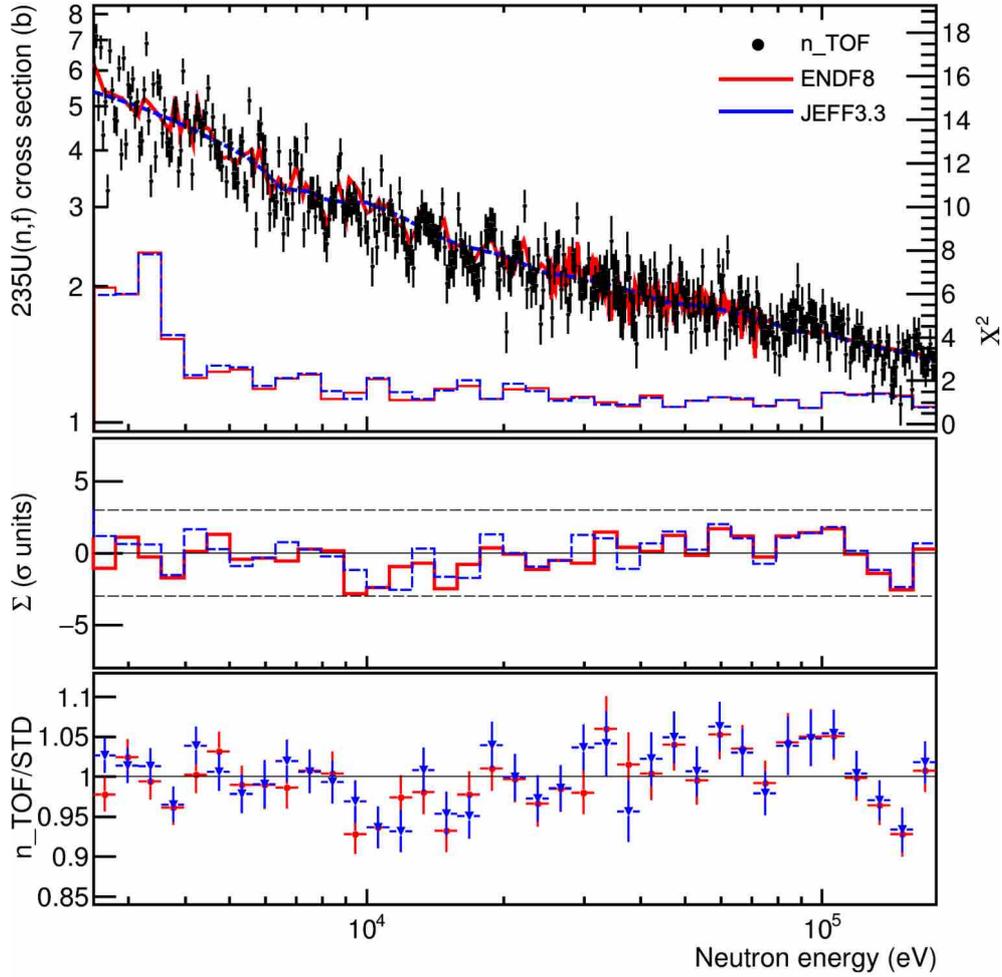

Figure 18 - (Color online) $^{235}$U(n,f) measured cross section of this work, in the 2-100 keV neutron energy range (with $\chi^2$, $\Sigma$ and ratio), compared to the ENDF-B/VIII and JEFF3.3 evaluations.

A check of the measured cross section was performed by averaging it in several energy intervals, and comparing the resulting values to the corresponding average values obtained by linear interpolation and integration between the recommended boundaries as indicated in [8] (GMAP output of point-wise cross sections), as well as to the ENDF/B-VIII and JEFF 3.3 evaluations, averaged in the same way in the same energy intervals. We remark that the boundary between the GMA nodes at 9.5 and 15 keV was modified with respect to ref.[8] from 12.25 to 10 keV, according to ref.[34]. The resulting values for the n_TOF data and the corresponding IAEA ones are listed in Table 2. The ratios and the corresponding normalized deviation $\Sigma$ are reported in Figure 19, along with the values for ENDF-B/VIII and JEFF3.3. A significant difference can be observed in at least three intervals with respect to the libraries, possibly indicating the need of further investigations by the evaluators.

Table 2 – Average cross section values by IAEA, and corresponding values from n_TOF, in the intervals defined by the GMA nodes [8]. The normalized deviation $\Sigma$ and the ratio are also listed.

| En [keV] GMA node | range [keV] | IAEA c.s. [b] | unc. Δ [b] | n_TOF c.s. [b] | stat. unc. [b] | n_TOF /IAEA | stat. unc. | $\Sigma$ (n_TOF - IAEA) |
|---|---|---|---|---|---|---|---|---|
| 3.5 | 3÷4 | 4.8177 | 0.063 | 4.8022 | 0.068 | 0.997 | 0.014 | -0.2 |
| 4.5 | 4÷5 | 4.2946 | 0.056 | 4.3729 | 0.074 | 1.018 | 0.017 | 1.1 |
| 5.5 | 5÷6 | 3.8415 | 0.050 | 3.8211 | 0.083 | 0.995 | 0.022 | -0.2 |
| 6.5 | 6÷7 | 3.3695 | 0.044 | 3.2713 | 0.073 | 0.971 | 0.022 | -1.3 |
| 7.5 | 7÷8 | 3.2291 | 0.042 | 3.2842 | 0.080 | 1.017 | 0.025 | 0.7 |
| 8.5 | 8÷9 | 3.0626 | 0.040 | 3.0644 | 0.082 | 1.001 | 0.027 | 0.0 |
| 9.5 | 9÷10 | 3.1062 | 0.040 | 2.9703 | 0.084 | 0.956 | 0.027 | -1.6 |
| 15 | 10÷17.5 | 2.6826 | 0.035 | 2.5668 | 0.034 | 0.957 | 0.013 | -3.5 |

| 20 | 17.5÷22 | 2.3528 | 0.031 | 2.3257 | 0.048 | 0.989 | 0.020 | -0.6 |
| 24 | 22÷27 | 2.1688 | 0.028 | 2.1763 | 0.048 | 1.003 | 0.022 | 0.2 |
| 30 | 27÷37.5 | 2.0451 | 0.027 | 2.078 | 0.044 | 1.016 | 0.022 | 0.7 |
| 45 | 37.5÷50 | 1.8824 | 0.024 | 1.9012 | 0.040 | 1.010 | 0.021 | 0.5 |
| 55 | 50÷60 | 1.8114 | 0.024 | 1.8458 | 0.044 | 1.019 | 0.024 | 0.8 |
| 65 | 60÷70 | 1.7533 | 0.023 | 1.8282 | 0.045 | 1.043 | 0.026 | 1.7 |
| 75 | 70÷80 | 1.6806 | 0.022 | 1.6403 | 0.043 | 0.976 | 0.026 | -0.9 |
| 85 | 80÷90 | 1.6059 | 0.021 | 1.6853 | 0.059 | 1.049 | 0.037 | 1.3 |
| 95 | 90÷97.5 | 1.5813 | 0.021 | 1.6833 | 0.066 | 1.064 | 0.042 | 1.5 |
| 100 | 97.5÷110 | 1.5691 | 0.020 | 1.6256 | 0.046 | 1.036 | 0.029 | 1.2 |
| 120 | 110÷135 | 1.4991 | 0.019 | 1.4872 | 0.030 | 0.992 | 0.020 | -0.4 |
| 150 | 135÷160 | 1.4422 | 0.019 | 1.3628 | 0.033 | 0.945 | 0.023 | -2.4 |
| 170 | 160÷175 | 1.4073 | 0.020 | 1.4173 | 0.041 | 1.007 | 0.029 | 0.2 |

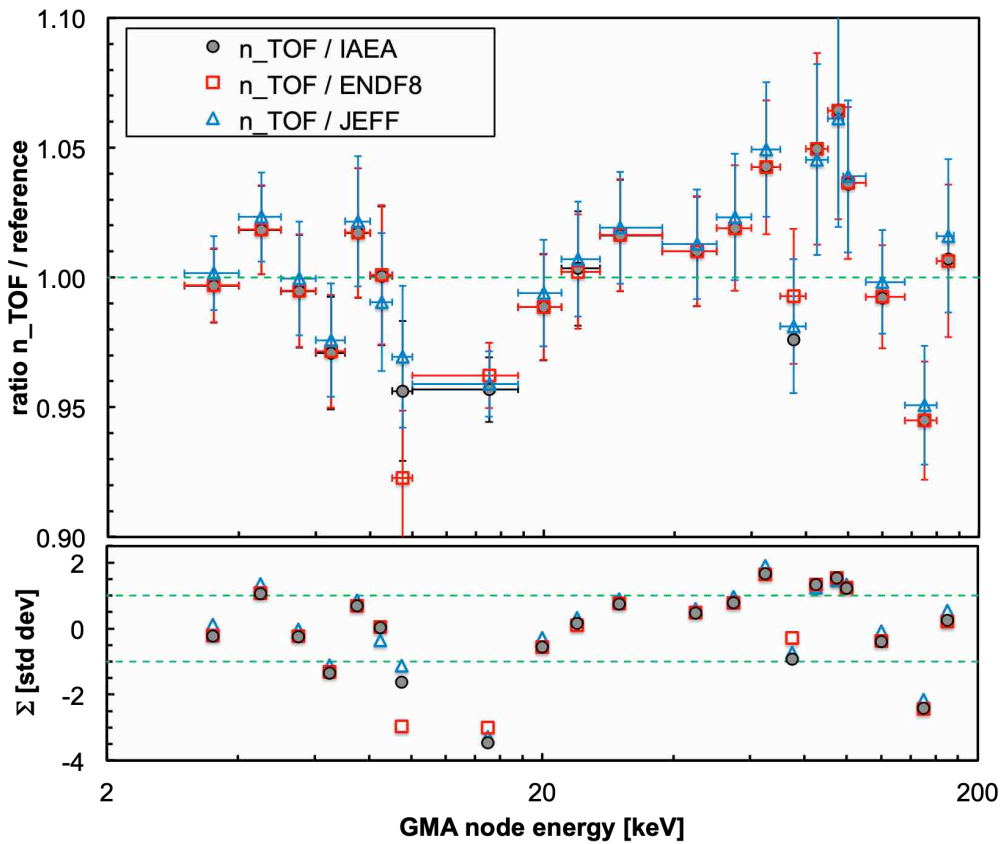

Figure 19 – (Color online) Top panel: ratio between the n_TOF average cross section in the GMA intervals recommended by IAEA and the corresponding values obtained from IAEA, ENDF-B/VIII, JEFF3.3. The horizontal bar indicate the averaging interval, the symbols are placed at the corresponding GMA node energy. Bottom panel: the normalized deviation Σ.

For a more quantitative assessment of the observed discrepancy, Figure 20 reports the ratio between the n_TOF cross section, integrated in a few relevant neutron energy regions, and the corresponding evaluated data. For completeness we included the evaluations from ENDF-B/VII, ENDF-B/VIII, JEFF3.3, JENDL4.0 and IAEA interpolation. The measured cross section, integrated between 9 and 30 keV, shows a deviation of 1.5÷2.5% relative to the evaluated data, pointing to a slight overestimate in the latter, as previously mentioned. An interesting conclusion can be drawn by splitting this energy range into two separate intervals, 9÷18 keV and 18÷30 keV and comparing the integrals with the corresponding integrals from the libraries. For the first integral a deviation up to 4.5% is observed (Σ<-3 for ENDF-B/VIII, JEFF3.3 and IAEA), and this is a significant indication that in that neutron energy range the cross section evaluation is overestimated and likely calls for a revision of these libraries.

The second integral (9÷18 keV) is in agreement with the evaluations within the statistical uncertainty. Between 30 and 100 keV the measured cross section is slightly larger than all the evaluations, whereas between 100 and 150 keV it agrees again within one standard deviation.

As a final remark, Figure 20 also shows that the cross section integrated in the highest measured neutron energy range (between 150 and 170 keV, i.e. in a region where the $^{235}$U(n,f) cross section is a standard), is in very good agreement with all the evaluations, further corroborating the robustness of the present results and conclusions.

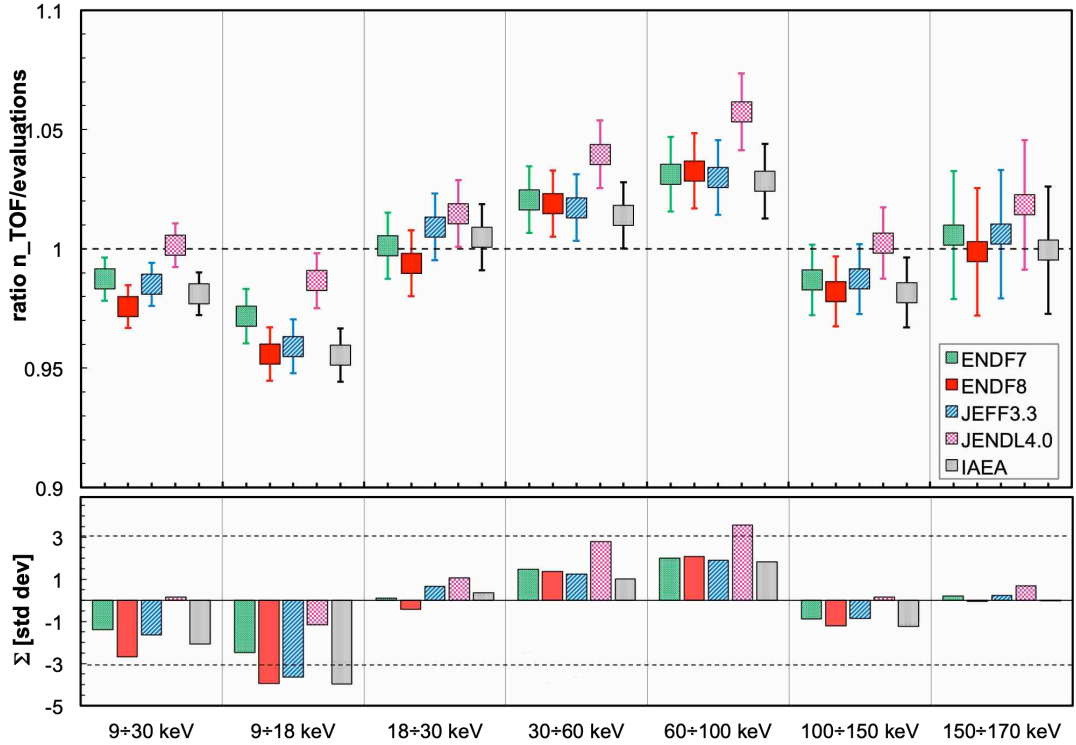

Figure 20 – (Color online) Top panel: ratio between the measured cross section, integrated in a few relevant intervals, and the corresponding values for the five reference libraries ENDF-B/VII, ENDF-B/VIII, JEFF3.3, JENDL4.0, IAEA. Bottom panel: the corresponding normalized deviation Σ (standard deviation units).

## 5 Conclusion

A high accuracy, high resolution measurement of the $^{235}$U(n,f) cross section, relative to the $^{6}$Li(n,t) and the $^{10}$B(n,α) reference reactions was performed at n_TOF in the Experimental Area 1, by means of stacks of samples and silicon detectors placed directly in the neutron beam. Data have been collected for all these three reactions in both forward and backward direction, to minimize the uncertainty related to angular anisotropy in the charged particle emission from the reference reactions. The measured count rate distributions as a function of neutron energy have been normalized to each other slightly above thermal neutron energy, in an interval recommended by IAEA. The new experimental setup and the employed analysis technique have resulted in an unprecedented low uncertainty of 1-2% on the data for all measured reactions.

The measured $^{6}$Li(n,t)/$^{10}$B(n,α) ratio of count rates, after applying all the relevant corrections, was found in remarkably good agreement with the ratio of the standard cross sections up to 170 keV, providing high confidence in the reliability of the experimental technique and results. A $^{235}$U(n,f) cross section was extracted relative to the weighted average of the two reference reactions from thermal energy to 170 keV. Up to 10 eV the measured cross section is in good agreement with the evaluated cross sections, while statistically significant differences are observed in the resolved resonance region (whose detailed analysis will be the subject of a dedicated follow-up paper). A clear indication of several statistically significant structures in the cross section above 2.25 keV was found, where the existing libraries only report average resonance parameters. Finally, a possible overestimate of the cross section by the major evaluated data libraries in the neutron energy range between 9 and 18 keV was found with a high confidence level, thus confirming what several previous measurements seemed to indicate in the 10÷30 keV range and calling for a possible revision of the libraries. Current developments of the experimental setup, that will make it less sensitive to the intense γ-flash in the neutron beam, might allow in the near future to extend the range of the measurable cross-sections to several MeV.

**Acknowledgments**

This research was partially funded by the European Atomic Energy Community (Euratom) Seventh Framework Programme No. FP7/2007-2011 under Project CHANDA (Grant No. 605203).

We also acknowledge the support of:
- the Narodowe Centrum Nauki (NCN), under Grant No. UMO-2016/22/M/ST2/00183;
- the Spanish Ministerio de Ciencia e Innovaciòn under grants FPA2014-52823-C2-1-P, FPA2017-83946-C2-1-P and the program Severo Ochoa (SEV-2014-0398);
- the Croatian Science Foundation under the project IP-2018-01-8570.